\documentclass[a4paper,12pt]{article}
\RequirePackage[OT1]{fontenc}
\RequirePackage{graphicx}    
\RequirePackage[english]{babel}    
\RequirePackage{natbib}
\RequirePackage[colorlinks,citecolor=blue,urlcolor=blue]{hyperref}
\RequirePackage{amsmath}
\RequirePackage{mathrsfs,amssymb}
\RequirePackage{amsfonts}
\RequirePackage{amstext}
\RequirePackage{multirow}

\begin{document}
\newcounter{trick}
\setcounter{trick}{\value{footnote}}
\renewcommand{\thefootnote}{\alph{footnote}}
\renewcommand{\thefootnote}{\arabic{footnote}}
\setcounter{footnote}{\value{trick}}

\begin{titlepage}
\title{Income Tax Evasion Dynamics: Evidence from an Agent-based Econophysics Model}


\author{Michael Pickhardt\footnote{Brandenburg University of Technology Cottbus, Postbox 10 13 44, 03013 Cottbus/Germany, Faculty 3, Institute of Economics, email: michael@pickhardt.com} and Goetz Seibold\footnote{Brandenburg University of Technology Cottbus, Postbox 10 13 44, 03013 Cottbus/Germany, Faculty 1, Institute of Physics and Chemistry, email: seibold@tu-cottbus.de}}

\date{November 29, 2011}

\maketitle


\begin{abstract}  
We analyze income tax evasion dynamics in a standard model of statistical mechanics, the Ising model of ferromagnetism. However, in contrast to previous research, we use an inhomogeneous multi-dimensional Ising model where the local degrees of freedom (agents) are subject to a specific social temperature and coupled to external fields which govern their social behavior. This new modeling frame allows for analyzing large societies of four different and interacting agent types. As a second novelty, our model may reproduce results from agent-based models that incorporate standard Allingham and Sandmo tax evasion features as well as results from existing two-dimensional Ising based tax evasion models. We then use our model for analyzing income tax evasion dynamics under different enforcement scenarios and point to some policy implications.     

\vspace*{2cm}


\vspace*{1cm}

\end{abstract}

\end{titlepage}

\section{Introduction}
\label{sec:Intro}

Agent-based tax evasion models have gained much popularity over recent years because they allow for analyzing tax compliance behavior in large populations of heterogeneous agents that interact with each other in a direct manner (see e.g.  \cite{koro07}). Moreover, these models allow for a high degree of complexity. For example, by simultaneously incorporating various policy parameters of the government, by endowing each individual agent with a different set of attributes regarding income, risk aversion, etc.  or by calibrating individual agent behavior with a diversity of different individual human behavior patterns, which may have been discovered in tax evasion experiments with human subjects, in fields such as economic psychology or by empirical analysis (see e.g. \cite{alm92}, \cite{and98}, \cite{kirch07}). To this extent, and in contrast to traditional models, agent-based tax evasion models allow for analyzing tax evasion dynamics in a fairly realistic way, which in turn may lead to new insights and policy options for combating tax evasion.  
  
In this paper, we develop and analyze an agent-based tax evasion model that is based on a standard model of statistical mechanics, the Ising model of ferromagnetism. Among other things, the model allows for the numerical simulation of tax evasion dynamics in very large populations of heterogeneous agents, where heterogeneity refers to several different behavioral patterns. We show that our model allows for reproducing previously published results obtained from fundamentally different types of agent-based models. We then use the model for analyzing tax evasion dynamics that follow from alternative enforcement scenarios to combat tax evasion. For example, we find that (i) for given parameter values there is a certain threshold for efficient audits, (ii) real world audit rates may well be too low to effectively curb tax evasion and (iii) that at some level direct agent interaction may effectively be a substitute for monetary penalty payments. 

The paper is organized as follows. In section two we provide some background on standard income tax evasion theory and existing agent-based tax evasion models. In section three, we develop the econophysics model of tax evasion and apply it to an analysis of audit efficiency and for some replication studies. Concluding remarks are provided in the final section.

\section{Background}
\label{sec:Back}

In this section we briefly compare and contrast the standard approach to income tax evasion with essential features of agent-based tax evasion models. Next, we review a few of these models and offer some background on the Ising model, which we use in section three for constructing a novel agent-based tax evasion model.

\subsection{Modeling Income Tax Evasion}

The neoclassical standard approach to income tax evasion considers a representative, self-reporting agent who declares an income $X$ so as to maximize expected utility, $EU$, according to, 
\begin{equation}
\label{AS1}
EU = (1-p) (W - \theta X) + p(W - \theta X - \pi (W - X)),
\end{equation}
where $W$ denotes the true income of the representative agent, $\theta$ is the tax rate on declared income and $\pi$ is the tax rate on undeclared income, where $\theta < \pi$ indicates a monetary penalty on income tax evasion and $p$ is the audit probability, with $0 \leq  p \leq  1$ [\cite{alling72}]. Given this modeling frame, risk neutral tax payers (i.e. (\ref{AS1}) is a linear function) declare their full income if $({\theta}/{\pi}) < p$, but declare nothing at all if $({\theta}/{\pi}) > p$. In contrast, risk-averse tax payers (i.e. (\ref{AS1}) is a concave function) may declare their income fully, partly or not at all. Other things being equal, the more risk-averse a tax payer is, the more compliant a tax-payer will be and both absolute and relative risk-aversion may play a role for the extent of tax evasion. Hence, in the standard income tax evasion model risk aversion is the driving force that allows for interior solutions. Of course, subsequent literature has developed various extensions and alternatives. For example, all-or-nothing decisions of risk-neutral tax payers may be avoided, if the audit probability is an increasing function of the amount of undeclared income [\cite{yitz87}], or if there are two or more income sources each having a different audit probability. Further, the penalty may be proportional to unpaid taxes [\cite{yitz74}], which ensures that the tax rate $\theta$ has no influence on the extent of tax evasion.

In any case, essential features of the neoclassical standard approach to tax evasion are: (i) each agent has perfect knowledge about his own expected utility function and the relevant parameter values, which allows him to maximize his own expected utility, (ii) the mathematical specification of the utility function leads to situations where the agent maximizes his expected utility either by an all-or-nothing decision or by an interior solution in which the agent may declare just some part of his true overall income, (iii) heterogeneity in agent behavior may be introduced by individualizing one or more parameter values that enter an agent's expected utility function, (iv) dynamics in the behavior of agents can be due to parameter changes only, (v) any kind of direct interaction among the agents is ruled out.
 
Agent-based tax evasion models deviate from the neoclassical standard approach in at least three ways. A feature that distinguishes any agent-based tax evasion model from the neoclassical approach is the direct interaction among agents. In particular, the behavior of all or at least some agents depends on the behavior shown by a well specified subgroup of other agents, say neighbors, within the same model. Besides this type of interaction is non-market based. Another important difference is that some agents, if not all agents, may not possess an utility function. Finally, in agent-based models tax evasion dynamics may be triggered by either parameter changes or by stochastic processes or a combination of both.

Although agent-based tax evasion models are a comparatively new tool for analyzing tax compliance issues, substantial differences already exist between these models or model types. Therefore, in the next subsection we briefly review the literature on agent-based tax evasion models.

\subsection{Agent-based Tax Evasion Models}
\label{BackTax}

As noted, the essential feature of any agent-based model is the direct non-market based interaction of agents, which is combined with some process that allows for changes in individual behavior patterns. Therefore, agent-based tax evasion models may be categorized according to the features of this individual interaction process. In fact, according to this criterion models may either fall into the economics domain or into the econophysics domain.
In the latter category, this process is driven by statistical mechanics using a model structure that is well known in physics, the Ising model. Examples include \cite{zak08,zak09}, \cite{lz08}, and \cite{lima10}. In contrast, if the interacting process is driven by parameter changes that induce behavioral changes via an utility function and/or by stochastic processes that are not related to models of statistical mechanics, these models belong to the economics domain. Examples include \cite{mp00}, \cite{davis03}, \cite{bloom04,bloom11}, \cite{koro07}, \cite{antun07}, \cite{szabo08,szabo09}, \cite{hopi10}, \cite{meder10}, \cite{NoZa} and \cite{Pelli11}.

\cite{bloom06} offers a review of the first three models and, therefore, in the following we just consider the remaining models. \cite{koro07} set up a model where agents compare expected payoffs from three alternatives: non-compliance, partial-compliance and full-compliance. The key feature of their model is the way in which audits and risk aversion are modeled. Agents never know the true audit probability. Rather, they are heterogeneous with respect to the perceived chance of being audited and they update their individual rate by observing how many members of their social network were actually audited. Likewise, agents are endowed with a different perceived taste for risk which they update based on the last period's interaction with the tax authority and by observing relative payoffs within their social network. Another important feature of their model is the social network size of nine members, of which eight are neighbors of the agent under consideration. Finally, they run a setting with no social network and three scenarios that differ with respect to the weight agents place on their neighbor's payoffs in updating their own perceived audit and risk parameters. With respect to the results they focus on stability of equilibria under changing enforcement regimes. Among other things, they find that even low rates of enforcement may support a highly compliant society of taxpayers, despite strong dynamics at the agent level.

\cite{szabo08,szabo09} model an entire economy with one industry consisting of several competing firms that need to hire employees, if they wish to produce output. Firms wish to minimize the expected effective salary, whereas employees want to maximize expected effective net income. Moreover, there is a government that provides public goods to the firms and the employees and there is a tax authority that conducts audits. To survive in the competitive environment, employers and employees can agree on various forms of contracts of which some are legal and others represent different forms of black labor, inducing tax evasion. Activities of the government, the tax authorities, or the social networks of employers and employees may alter the dynamics of tax compliance in one way or another. 

\cite{meder10} consider a number of tax evasion models of which one is agent-based. In this model tax payers have identical utility functions and can make binary decisions, to comply or to evade. Relevant for this decision is the level of individual utility, which in turn depends on a combination of internal and external factors. The internal factor is utility that depends on the provision of a public good and the external factor is observed behavior within the social network of an agent. Moreover, a stochastic process is added with a fixed probability of random switches between the two states, compliance and evasion. However, no audits or penalties are considered. Among other things, they find that efficiency in the provision of public goods reduces tax evasion. 

\cite{antun07} and \cite{hopi10} both consider a model with different behavioral types of which one type more or less complies with a rational tax payer of the Allingham and Sandmo type. They then consider income tax evasion dynamics that results from alternative governmental policies and also include lapse of time effects (back auditing). In fact, the model of \cite{hopi10} may be considered as the agent-based model that is closest to the \cite{alling72} model. In particular, with 100 percent selfish a-type agents (cf. below) the Hokamp and Pickhardt model represents a numerical version of the \cite{alling72} model.

\cite{NoZa} deal with endogenous norm formation over the life cycle of agents. In particular, their simulations support the view that older people hold stronger moral attitudes (i.e. evade less taxes) than younger agents due to an age effect rather than a cohort effect. In their model both personal and social norms influence an agent's decision of whether to evade taxes or not. Moreover, they allow for heterogenous norm-updating via several social psychology mechanisms (cognitive dissonance, self-signaling, and conformity with social network preferences). Otherwise, they follow the segregation model approach of \cite{sch71}. 

\cite{Pelli11} consider a model with heterogenous agents who differ with respect to income, risk-aversion, preferences for public expenditure, and believes about the actual level of tax compliance in the population. Audits and interactions among the agents are randomized and the government not only sets tax and penalty rates, but also spends all revenue to finance public expenditure. The population is set to 1,000 agents, each having a Cobb-Douglas utility function with their after tax monetary income, their perceived value of public expenditure and some relevant individual characteristics as arguments. Findings include that equilibrium situations still arise even in the heterogeneous case, that the presence of public expenditure may establish a positive relationship between the tax rate and tax evasion without assuming tax morale or the like, and that individual characteristics of the agents matter more for individual compliance than audit policy parameters. However, at least the latter finding may depend on the fact that their model does not allow for auditing past periods (back auditing).   

Agent-based tax evasion models that fall into the econophysics domain differ fundamentally from the aforementioned models of the economics domain. An important difference is that in an econophysics model none of the agents has an utility function of some sort. Rather, agents are in one of two possible states 'evading' or 'compliant' and the transition between both is a stochastic process. The latter is influenced by the behavior of neighbors and a global influence denoted as 'temperature', which has, just like a public good, a nonrival and simultaneous impact on the extent to which all agents are influenced by their respective neighbors. In fact, this difference has a number of implications and consequences. First of all, econophysics models are not based on the individualistic premises on which mainstream neoclassical economics rests. Therefore, in these models no agent possesses individual properties such as age, income, risk preferences, etc. A major consequence of this feature is that in econophysics models no monetary penalties can be charged and penalties cannot be differentiated with respect to the individual behavior or individual properties of the evader. Rather, if caught, the penalty of an evader in an econophysics model consists of an obligation to be compliant for a specified number of periods. To this extent, penalties in econophysics models differ fundamentally from those charged in economics models. Another important consequence is that econophysics models cannot consider any past decisions of tax payers. Moreover, agents in econophysics models are always limited to a binary decision space, in the present context they can be either compliant tax payers or evaders. Hence, the rate of tax evasion is measured in these models as the ratio of evaders over the total number of agents.

Yet, despite these fundamental differences an interesting aspect is that on the aggregate level econophysics models may mimic results obtained by economics models and that both model types may lead to stable equilibrium situations (steady states), which are characterized by a prevailing positive rate of tax evasion. Essentially, this implies that the results of individual rational behavior patterns may be reconstructed at the macro or aggregate level by means of natural stochastic processes such as statistical mechanics. In addition, econophysics models generate some interesting policy conclusions. For example, \cite{zak08} find that even very small levels of enforcement are sufficient to establish almost full tax compliance and \cite{zak09} conclude that regardless of how strong group influence may be, enforcement always works to enhance tax compliance. However, in both cases the adjustment process may take a very long time.

\vspace{1.0cm}

\subsection{Statistical Mechanics: The Ising Model}
\label{BackIsing}

\begin{figure}[htb] \begin {center} \includegraphics[width=12cm,clip=true]{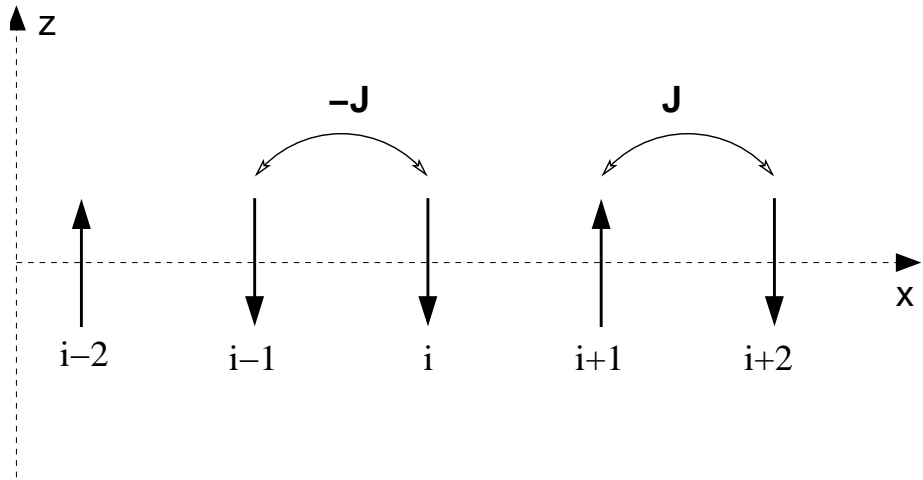} \caption{Sketch of the one-dimensional Ising model. Spin vectors are placed on discrete sites $i$ on a one-dimensional chain. The spin variables can take the value $S_i=+1 (-1)$ when the vector points in the positive (negative)  $z$-direction. The interaction energy between parallel spins is $E=-J$ and $E=+J$ between antiparallel spins.} \label{figising} \end {center}\end{figure}

The Ising model was originally introduced by the German physicist Wilhelm Lenz in order to study the formation of ferromagnetism. Lenz gave it as a problem to his student Ernst Ising who presented the solution for the one-dimensional system in his PhD thesis [\cite{ising25,brush67}]. It can be viewed as the simplest model for interacting magnetic moments (here and in the following referred to as 'spins'), which only can take the values $\pm 1$. As illustrated in Fig. \ref{figising}, the spins
can thus be represented as vectors $S_i$ which are attached to a lattice point (Fig. \ref{figising} shows a one-dimensional example). These vectors are only allowed to point in the positive ($S_i=+1$) or negative ($S_i=-1$) z-direction. The interaction energy $E_{int}$
\begin{equation}\label{eq:eint} E_{int}=-\sum_{ij}J_{ij}S_i S_j \end{equation}
drives the orientation of the spins where in the simplest case the interaction constant $J_{ij}\equiv J$ is only non-zero between adjacent sites $i$ and $j$. As can be seen from Fig. \ref{figising}, the corresponding interaction energy is $E_{int}=-J$ for parallel and $E_{int}=J$ for antiparallel spins. A positive (negative) exchange constant $J>0$ thus favors  parallel (antiparallel) alignment of the spins since the configurations are associated with a lower energy.

However, the spin configuration of the Ising model also depends on temperature $T$, which tries to disorder the spins. The associated characteristic quantity is the thermal energy $E_{thermal}=kT$, where $k\approx 1.38 \cdot 10^{-23} m^2 kg/(s^2 K)$ is the Boltzmann constant which depends on units $m$ (meter), $kg$ (kilogramm), $s$ (seconds), and $K$ (Kelvin).  Basically the system is in a (dis)ordered state when the interaction energy per spin is much (larger) lower than the thermal energy. A special case in this regard is the one-dimensional Ising model which does not have a phase transition at finite temperatures, but only is in an ordered state at exactly $T=0$. The situation is different for higher dimensional lattices. \cite{onsager44} analytically demonstrated the occurrence of a finite temperature  order-disorder transition on a two-dimensional square lattice at $kT_c\approx 2.269J$, where $T_c$ denotes the critical temperature and $J$ is the exchange constant between nearest-neighbor sites as defined in Eq. \ref{eq:eint}.  

In contrast, for dimensions $d>2$ despite its simplicity the model can only be solved numerically. For example, numerical evaluation of the transition temperature for a three-dimensional cubic lattice yields $kT_c \approx 4.51 J$, below which the spins are in an ordered state. Note that a spin on a $d$-dimensional hypercubic lattice is attached to $2d$ nearest neighbors so that the higher the dimension (or, more precisely, the coordination number) the thermal energy has to overcome  a larger interaction energy to disorder the spins. In order to simplify notation we will measure in the following temperature in units of energy (i.e. we set $kT \to T$). In section three we will also couple the spins to an external magnetic field $B_i$, which can be different at every lattice site. These fields tend to align the magnetic moments of the spins, which can be described by the energy contribution $E_{B}=-S_i B_i$. Thus, a positive field $B_i > 0$ lowers the energy for a positive spin value $S_i=+1$, whereas a negative field favors negative spins $S_i=-1$. 

Nowadays, the Ising model is one of the standard models in statistical physics. Moreover, it has become popular in a large variety of different fields like biology, sociology, economics etc. because it is one of the simplest models that describes the interaction of entities which try to adjust their behavior in order to be conform with that of their  neighbors. For example, Thomas C. Schelling's (1971) segregation model roughly corresponds to an Ising model at $T = 0$ as shown by \cite{stauffer08}, \cite{stauffer07} and \cite{mueller08}.

\section{Modeling Income Tax Evasion Dynamics}
\label{sec:Model}

In this section we model tax evasion dynamics by developing an agent-based econophysics model. In section \ref{sec:ModelTwo} we first introduce the standard two-dimensional Ising model and reproduce some selected results of Zaklan et al. (2009). Next, we extend this model by incorporating features of heterogeneous multi-agent simulations. We apply the model to an analysis of different enforcement scenarios and for replicating some results of \cite{hopi10}. All results presented in this section are obtained from a Fortran code, which is available upon request.

\begin{figure}[htb] \begin{center} \includegraphics[width=12cm,clip=true]{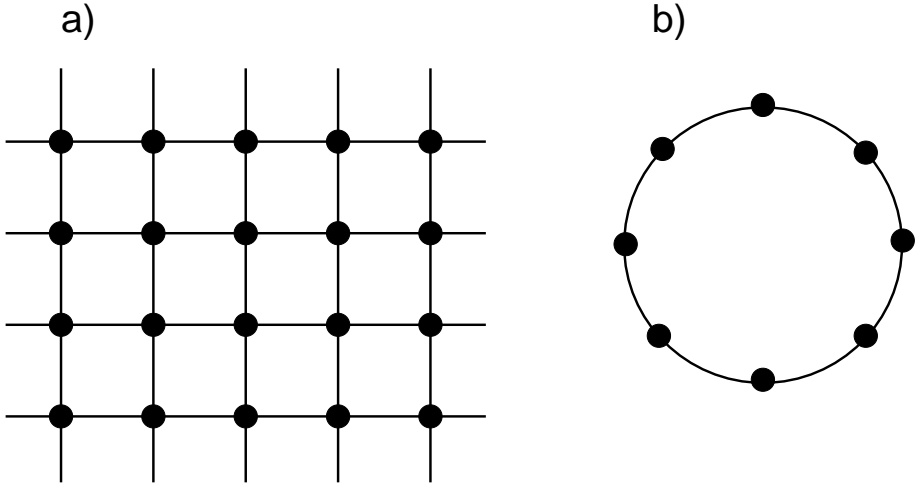} \end{center} \caption{(a) Square lattice considered in Sec. \ref{sec:Model}. The spin degrees of freedom sit on the lattice sites marked by solid dots and each spin interacts with its four adjacent neighbors. (b) Sketch of a ringworld structure as additionally considered in  Sec. \ref{sec:HOPI}. Here each spin is coupled to two adjacent neighbors only.} \label{fig5} \end{figure}

\subsection{The Two-Dimensional Ising Model}
\label{sec:ModelTwo}
Our considerations are based on the two-dimensional Ising model Eq. (\ref{eq:eint}), implemented on a $1000 \times 1000$ square lattice (see Fig. \ref{fig5}).\footnote{For example, \cite{zak09,zak08} and \cite{lz08} consider, in addition, 
alternative lattice structures like the scale-free Barab\'{a}si-Albert network or the Voronoi-Delaunay network. In addition \cite{lima10} considers Erd\"os-R\'{e}nyi random graphs and finds that the results for these alternative lattices {\it ceteris paribus} do not differ fundamentally from those obtained with a square lattice.} In addition to the 'bare' model Eq. (\ref{eq:eint}) we now allow for the coupling of each spin to a local magnetic field $B_i$. The corresponding hamiltonian reads as,
\begin{equation}\label{eq:ising} H=-\sum_{ij} J_{ij} S_i S_j - \sum_i B_i S_i \end{equation} 
The spin degrees of freedom $S_i$ can take the values $S_i=\pm 1$ and  $J_{ij}$ denotes the interaction between two spins on sites $i$ and $j$. We take $J_{ij}\equiv J>0$ to be a constant within a given 'interaction range' (corresponding to the nearest neighbors, i.e. four agents in the present section) and  $J_{ij}\equiv 0$ otherwise. In the present context $S_i=1$ is interpreted as a compliant tax payer and $S_i=-1$ as a non-compliant one. We use the heat-bath algorithm [cf. \cite{krauth}] in order to evaluate statistical averages of the model. The probability for a spin at lattice site $i$ to take the values $S_i=\pm 1$ is given by
\begin{equation}\label{eq:prob} p_i(S_i)=\frac{1}{1+\exp\{-[E(-S_i)-E(S_i)]/T\}} \end{equation}and $E(-S_i)-E(S_i)$ is the energy change for a spin-flip at site $i$. 
Upon picking a random number $0 \le r \le 1$ the spin takes the value
$S_i=1$ when $r < p_i(S_i=1)$ and $S_i=-1$ otherwise.
One time step then corresponds to a complete sweep through the lattice. In analogy to \cite{zak09} we further implement the probability $p_a$ of a tax audit. If tax evasion is detected the agent has to remain compliant over $h$ periods of time. \cite{zak09,zak08} and \cite{lz08} interpret this number of $h$ periods of enforced compliance as a penalty or as a consequence of shame and guilt feelings, respectively. But, as noted, this kind of penalty is fundamentally different from monetary penalties applied in tax evasion models of the Allingham and Sandmo type. Rather, periods of after audit compliance are used elsewhere to accommodate extreme subjective audit expectations (see e.g. \cite{hopi10}).

\subsubsection{Zero-field Results}
For illustrative purposes and in order to define the limits  for our multi-agent generalization below, we consider first the zero-field case ($B_i=0$ in Eq. \ref{eq:ising}). This corresponds to the model introduced by Zaklan et al. (2009, 2008) where the system is composed of one type of agent only and spin-flip probability is solely determined by the parameter $J/T$. In the present context, the ratio $J/T$ can be viewed as a  measure for the autonomous behavior of the tax payers. In the high-temperature limit $J/T <<1$ an agent at site $i$ decides autonomously from his or her social environment (i.e. $p_i\approx 1/2$), whereas for $J/T \gtrsim 1$ the probability $p_i$ strongly depends on the compliance of his or her neighbors within the social network. Note that this is related to the ferromagnetic phase transition of the two-dimensional Ising model (for $B_i=0$), which occurs at $J/T_c \approx 1/2.269$ as discussed in the previous section. In the following we consider a population of $1.000.000$ agents, except in Sec. \ref{sec:HOPI} where the population is set to $150.000$ agents.

\begin{figure}[htb] \begin{center} \includegraphics[width=12cm,clip=true]{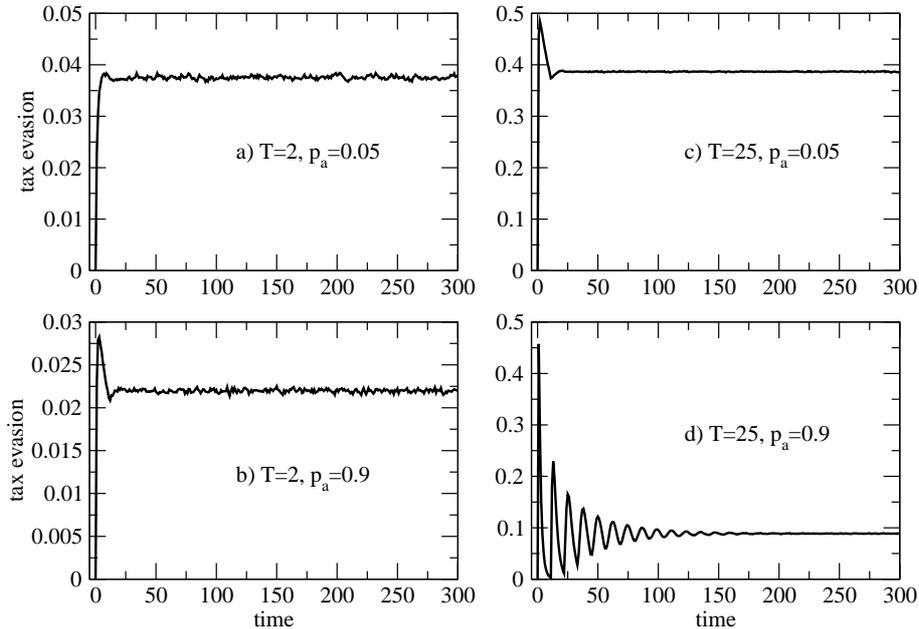} \end{center} \caption{Tax evasion dynamics for $B=0$. Panels a and b display results for agents with a social temperature of $T=2$ and audit probabilities $p_a=0.05$ and $p_a=0.9$, respectively. Panels c and d show the corresponding results  for a social temperature of $T=25$. Reference energy scale is $J\equiv 1$. Tax audits enforce compliance over $h=10$ time steps.} \label{fig1a} \end{figure}

Fig. \ref{fig1a} (panels a and b) displays the corresponding extent of tax evasion (i.e. the fraction of $-1$-spins) as a function of time for small and large audit probabilities $p_a=0.05$ and $p_a=0.9$, respectively. We define $J\equiv 1$ as reference energy scale and as initial condition we set all agents to 'compliant', i.e. $S_i=1$. Due to the small 'social temperature' $T_i=2$ (which is, thus, of the same order than the exchange energy) only a small number of agents become non-compliant in the first time step and because the audit probability is also small, tax evasion $p_{te}$ increases continuously up to some saturation value $p_{te}\approx 0.038$. At this value the system of agents has reached equilibrium between the influence of the neighborhood to become a non-compliant tax payer and the tax audits which enforce compliance over $h=10$ time periods. Upon increasing the audit probability to $p_a=0.9$ (panel b) the number of tax evaders reaches its maximum already after about three time steps and then starts to decrease due to the high audit probability. After $10$ time steps, initially detected tax evaders may again switch from compliant to non-compliant behavior, which leads to the observed small increase of $p_{te}$ before it reaches its saturation value of $p_{te}\approx 0.022$. Note that the 'social temperature' $T_i=2$ is below the ordering temperature $T_c=2.269$ of the two-dimensional Ising model. For zero audit probability the long-term tax evasion would therefore also converge to a small value ($p_{te}\approx 0.045$), corresponding to the fraction of 'minority spins' at this particular temperature. The effect of a finite audit probability is, therefore, just a further reduction from this small value.

In contrast, above the ordering temperature, the zero audit equilibrium state is of course reached for an equal number of compliant and non-compliant tax payers, i.e. $p_{te}=0.5$. Finite audit probabilities then lead to a further reduction of this equilibrium value. This situation is depicted in panels c and d of Fig. \ref{fig1a}, which display the extent of tax evasion for a system build up from agents who decide mainly autonomously, due to the large 'social temperature' $T=25$ (i.e. much larger than the exchange energy). As a consequence, already after two time steps approximately half of the agents become non-compliant which in turn induces a reduction of $p_{te}$ due to the auditing. It is remarkable that even for small audit probability $p_a=0.05$ (Fig. \ref{fig1a}c) $p_{te}$ decreases by $\approx 20\%$ before it saturates at $p_{te}\approx 0.39$ after more than $10$ time steps. Upon invoking a larger audit probability, $p_a=0.9$, the number of autonomous agents decreases rapidly after the first initial time steps and consequently $p_{te}$ decreases to almost zero within the first $10$ time steps. Half the fraction of those who have been detected at the first time step will then select the possibility of non-compliance again, which leads to the oscillatory behavior of $p_{te}$. A stable situation is only reached at $>> 150$ time steps at $p_{te}\approx 0.09$.

Note that the results of Fig. \ref{fig1a}, which we have discussed so far,  correspond to those presented by  \cite{zak09} [Fig. 1 and Fig. 4]. In fact, our results represent the first independent replication of results originally obtained from an agent-based econophysics tax evasion model.

\subsubsection{Finite Fields Results}
We now follow the scheme proposed in \cite{lz08} and discuss
the case of societies where the agents have an endogenous preference with regard to compliance or non-compliance. This feature can be incorporated via the field parameter $B_i$ in  Eq. \ref{eq:ising}. A value of $B_i < 0$ enhances the probability of $S_i=-1$ values and, therefore, corresponds to selfish agents, who try to maximize their payoff by non-compliant behavior. On the other hand, a positive field $B_i >0$ favors positive values $S_i=+1$ and, therefore, can be used to describe ethical agents who wish to be compliant, but may make occasional mistakes \footnote{\cite{lz08} interpret a negative (positive) field as an agent's low (high) confidence in governmental institutions.}. With regard to the Ising model Eq. \ref{eq:ising}, a dominant endogenous behavior further requires $|B_i| >> J$ in both cases.  The results in the present section are obtained for a society of identical agents, i.e. $B_i\equiv B$ with $|B|/T=3$ and $J/T=0.2$. 

\begin{figure}[h!] \begin{center} \includegraphics[width=12cm,clip=true]{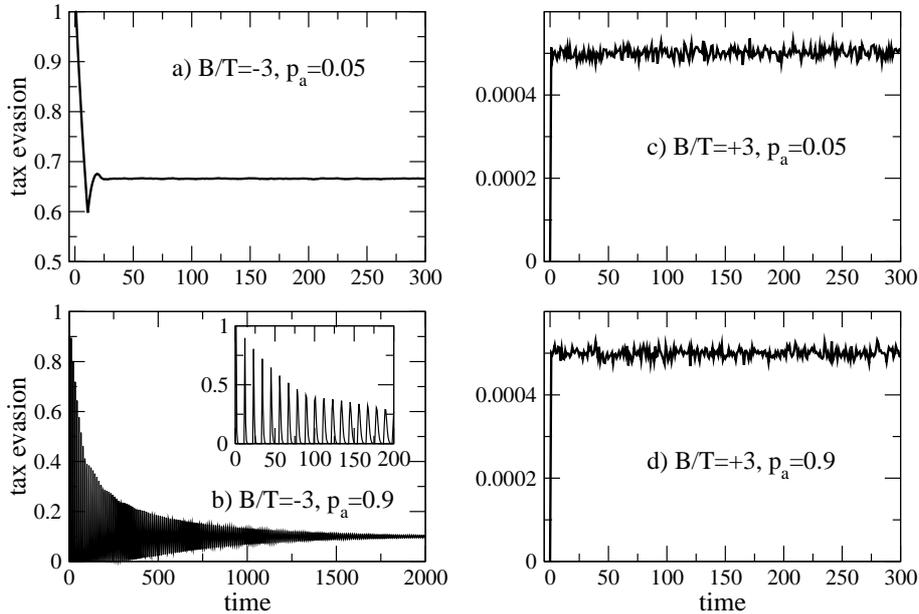} \end{center} \caption{Tax evasion dynamics for $B\ne 0$. Panels a and b display results for selfish agents, i.e. endogenous non-compliant agents with a bias towards tax evasion $B/T=-3$ and audit probabilities $p_a=0.05$ and $p_a=0.9$, respectively. The inset to panel b displays the small time behavior to visualize the oszillations in $p_{te}$. Panels c and d show the corresponding results for  ethical agents with a tax evasion bias $B/T=+3$ and audit probabilities $p_a=0.05$ and $p_a=0.9$, respectively. For all results the ratio between interaction constant $J$ and
temperature $T$ is set to $J/T=0.2$. Tax audits enforce compliance over $h=10$ time steps.}  \label{fig1b} \end{figure}

The first case of endogenous non-compliant selfish agents (i.e. $B/T=-3$) is shown in panels a and b of Fig. \ref{fig1b}. At time step zero we correspondingly set the share of tax evasion to $p_{te} =1$. For zero audit probability the equilibrium state would, therefore, yield a tax evasion rate close to one ($p_{te}\approx 0.9995$) due to rare spin flip events. However, a small audit probability $p_a=0.05$ reduces tax evasion significantly because at each time step five percent of the remaining non-compliant agents are forced to become compliant. After  $h=10$ time steps the first detected agents can become non-compliant again, which after a small increase leads to the stabilization of tax evasion at $p_{te}\approx 0.67$. A larger audit probability $p_a=0.9$ (Fig. \ref{fig1b}b) again drives tax evasion rapidly to zero and, in addition, induces  weaker peaks at those time steps (i.e. multiples of $h+1=11$) when 'forced' compliant agents have the possibility to be non-compliant again (cf. inset to Fig. \ref{fig1b}b). When the peak height is below some value (after $\approx 300$ time steps), then the audit is no longer able to reduce
tax evasion to $p_{te} \sim 0$. However, the minimum value starts to increase, leading again to an equilibrium state on very long time scales.

Finally, the time evolution for ethical agents (i.e. $B/T=3$)  is rather obvious. Initial tax evasion is set to $p_{te} =0$ and there is only a very small probability that one of the agents becomes non-compliant. This probability is controlled by the ratio $B/T=3$ and would become zero in the limit  $B/T \to \infty$. Also, since ethical agents wish to minimize tax evasion, the results are also almost independent of the audit probability. Hence, any positive audit probability would be inefficient in this case.

\subsection{Heterogenous Agents Simulations within the Two-dimensional Ising Model}
\label{sec:ModelMulti}
It is now straightforward to generalize the results of the previous section in order to describe societies which are characterized by behaviorally different types of agents. To achieve this, we define a local temperature $T_i$ at each site so that an ensemble of heterogeneous agents can be classified according to their ratios $J/T_i$ and $B_i/T_i$. As already outlined in the previous section the parameter $J/T_i$ describes the exogenous conditioning of agent $i$, whereas his or her endogenous (non-manipulable) code of conduct is measured by the ratio $B_i/T_i$.

Following \cite{hopi10}, we introduce the following four types of agents: (i) \textit{selfish a-type agents}, which take advantage from non-compliance and, thus, are characterized by $B_i/T_i < 0$ and $|B_i| > J$; (ii) \textit{copying b-type agents}, which copy tax behavior of their social environment or neighborhood. This can be modelled by $B_i << J$ and $J_i/T_i \gtrsim 1$; (iii) \textit{ethical c-type agents}, who are practically always compliant and which are parametrized by $B_i/T_i > 0$ and $|B_i| > J$; (iv) \textit{random d-type agents}, which act by chance, within a certain range, due to some confusion caused by tax law complexity. We implement this behavior by $B_i<<J$ and $J/T_i <<1$. Here and in the following all parameters are measured with respect to $J \equiv 1$. Within these definitions the results presented in the previous section correspond to homogeneous societies of b-type (Fig. \ref{fig1a}a,b), d-type (Fig. \ref{fig1a}c,d), a-type (Fig. \ref{fig1b}a,b), and c-type agents (Fig. \ref{fig1b}c,d). Note that by introducing b-type and d-type agents, we extend the model of \cite{lz08}, who consider just a-type and c-type agents.

\subsubsection{Enforcement Regimes and Tax Evasion Dynamics}

We now take the number of agents for each behavioral type as fixed, but introduce some heterogeneity for each type. This is achieved by randomizing the parameters $J_i/T_i$ and $B_i/T_i$ within a given range, which is specified in Table \ref{tab1}, and relevant parameters for the following results are given in Table \ref{tab2}.

\begin{table}
\caption{Parameter ranges for each group of agents. Note that $T$ and $B$
are measured in units of $J\equiv 1$.}
\label{tab1} 
\begin{center}
\begin{tabular}{|l|c|c|}\hline
Agent-type  & temperature & field \\ \hline\hline
a  & $T_{ac}$  & $-|B_a^{max}| < B_a < -|B_a^{min}|$ \\ \hline
b & $T_b^{min} < T_b < T_b^{max}$ & 0 \\ \hline
c & $T_{ac}$ & $|B_c^{min}| < B_c < |B_c^{max}|$ \\ \hline
d & $T_d^{min} < T_d < T_d^{max}$ & 0 \\ \hline
\end{tabular}
\end{center} 
\end{table}

\begin{table}
\caption{Parameter set for the results shown in Figs. \ref{fig2} to \ref{fig3}.
Note that $T$ and $B$
are measured in units of $J\equiv 1$.}
\label{tab2}
\begin{center}
\begin{tabular}{|l|l|l|l|l|l|l|l|l|} \hline
$T_{ac}$ & $T_b^{min}$ & $T_b^{max}$ & $T_d^{min}$ & $T_d^{max}$ & $B_a^{min}$ & 
$B_a^{max}$ & $B_c^{min}$ & $B_c^{max}$\\  \hline\hline
5 & 1 & 3 & 10 & 30 & 10 & 20 & 10 & 20 \\ \hline
\end{tabular}
\end{center} 
\end{table}

In order to interpret the tax evasion dynamics for the heterogeneous agents society, we report in Table \ref{tab3} the spin probabilities [cf. Eq. (\ref{eq:prob})] for each type of agent within the boundaries set by the values given in Tables \ref{tab1} and \ref{tab2}. Essentially, an a-type agent has a large tendency to display non-compliant behavior,  independent of its neighbors. As noted, b-type agents copy the behavior of neighbors and consequently compliant b-type agents have a small probability in a non-compliant neighborhood and \textit{vice versa}. Ethical c-type agents have a large probability to be compliant independent of the neighborhood. Finally, d-type agents have spin probabilities around $50\%$, which more or less weakly depend on the behavior within their neighborhood and their own individual behavior pattern.

\begin{table}
\caption{Dependence of spin probabilities $p_i(S_i)$ [cf. Eq. (\ref{eq:prob})] for different agent  types in a specific neighborhood $\sum_j S_j$, where $j$ denotes nearest neighbors of $i$. $S=\pm 1$ specifies whether the corresponding agent is compliant or  not and we report minimal / maximal $p_i$ according to the boundaries of the parameter distribution reported in tables \ref{tab1} and \ref{tab2}.}
\label{tab3}
\begin{center}
\begin{tabular}{|r|l|l|l|l|l|l|l|l|} \hline
\multirow{2}{*}{$\sum_j S_j$} & \multicolumn{2}{|c|}{a-types} &
\multicolumn{2}{|c|}{b-types} &
\multicolumn{2}{|c|}{c-types} &
\multicolumn{2}{|c|}{d-types} \\ \cline{2-9}
 & S=-1 & S=+1 & S=-1 & S=+1 & S=-1 & S=+1 & S=-1 & S=+1 \\ \hline \hline
-4 &99.6/100 &0 / 0.4 &69/93.5 &6.5/31 &0.2/8.3 &91.7/99.8 &56.6/69 & 31/43.4 \\ \hline
-2 &99.2/100 & 0/0.8&  59.9/79.1& 20.9/40.1 &0.1/3.9 & 96.1/99.9& 53.3/59.9&40.1/46.7  \\ \hline
0 &98.2/100 & 0/1.8& 50/50&  50/50 & 0/1.8& 98.2/100&50/50 & 50/50 \\ \hline
2 & 96.1/99.9& 0.1/3.9& 20.9/40.1& 59.9/79.1&0/0.8 & 99.2/100 &40.1/46.7 &53.3/59.9  \\ \hline
4 &91.7/99.8 & 0.2/8.3& 6.5/31& 69/93.5& 0/0.4&99.6/100 &31/43.4 & 56.6/69 \\ \hline
\end{tabular}
\end{center}
\end{table}

\begin{figure}[h!]
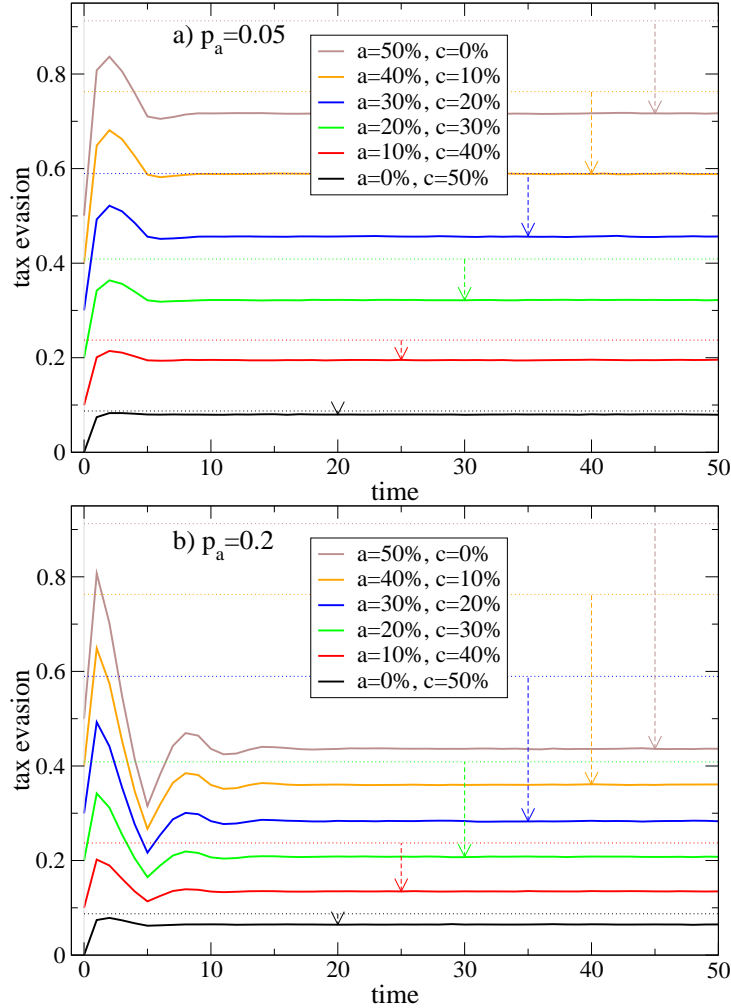

\begin{center}
\includegraphics[width=9.5cm,clip=true]{fig5a.eps}
\includegraphics[width=9.5cm,clip=true]{fig5b.eps}
\end{center}
\caption{Tax evasion dynamics for an ensemble of heterogeneous a- to d-type agents. The percentage of b- and d-type agents is fixed to $b=35$ percent and $d=15$ percent, respectively. The percentage of a- and c-type agents is varied according to the labels and the order within each label corresponds to the solid lines shown in each panel. Audit probability (a) $p_a=5\%$, (b) $p_a=20\%$ and $h=4$. The horizontal dotted line marks the stationary tax evasion value  without audit and the vertical arrows indicate the difference to the stationary value with audit.}
\label{fig2}
\end{figure}

In the results displayed in Fig. \ref{fig2} we fix the percentage of b- and d-type agents to $n_b=35\%$ and $n_d=15\%$, respectively (see Hokamp and Pickhardt, 2010) and vary the ratio of a- and c-type agents for two audit probabilities $p_a=0.05$ (panel a) and $p_a=0.2$ (panel b). The initial tax evasion is set to the fraction of a-type agents, i.e. $p_{te}=n_a$. 

Let us first discuss the case of a small audit probability and $n_a=0$ (first curve from below in Fig. \ref{fig2}a). Here the society consists of copying b-types ($35\%$), ethical c-types ($50\%$) and random d-type agents ($15\%$). The time evolution thus is similar to a mixture of the homogeneous agent systems  shown in Figs. \ref{fig1a}a,c and \ref{fig1b}c, i.e. the stationary state for large time steps is determined by non-compliant b- and d-type agents, the number of which is limited by the audit. Note, however, that the difference to the homogeneous cases lies in the fact that now the copying b-type agents are embedded in a 'matrix' of ethical c-type agents, which almost never change their behavior. One observes that the stationary value is not reduced significantly with respect to the limit without audit (indicated by the dotted horizontal lines), since the audit can essentially detect only non-compliant d-type agents. 

\begin{figure}[thb] \begin{center}\includegraphics[width=12cm,clip=true]{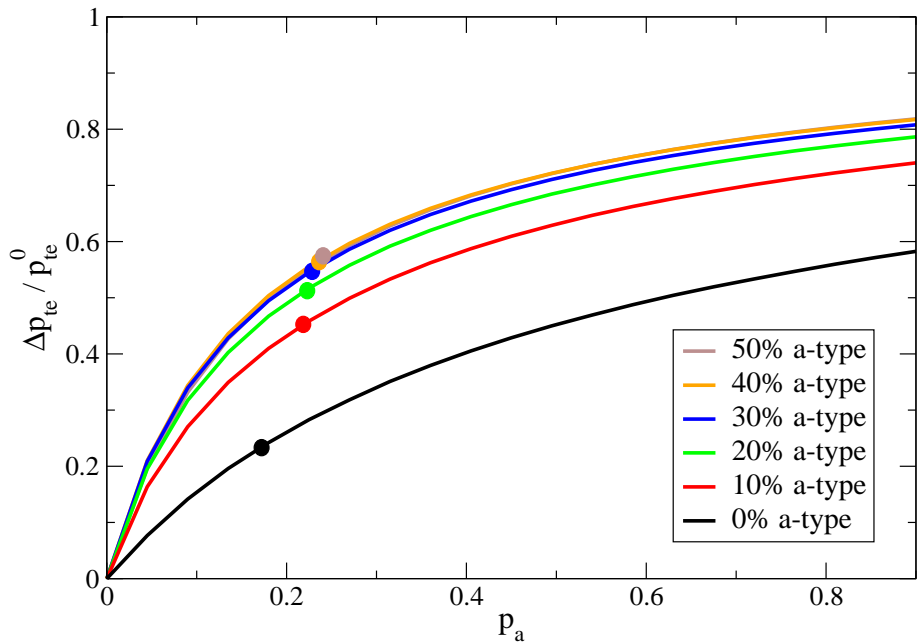}\end{center} \caption{Difference in tax evasion (stationary state) between zero $(p_{te}^0)$ and finite audit probability $\Delta p_{te}$ 
normalized to $p_{te}^0$ for different percentages of a-type agents. The circles indicate the point where the slope of the curves equals 'one'.}
\label{fig3a}
\end{figure}

The other limiting case is that for $n_a=50\%$ (first curve from above in Fig. \ref{fig2}a), where half of the society consists of tax evading a-type agents. During the first time step most b-type agents copy non-compliant behavior so that $p_{te}$ rises slightly above $80\%$. Moreover, a comparison of the stationary states with and without audit, $\Delta p_{te}$, reveals a significant reduction, because a-, b- and d-type agents may be forced to be compliant over a period of $h$ time steps. This reduction obviously increases for larger audit probabilities, as can be deduced from Fig. \ref{fig2}b. Naturally, a small audit probability mostly affects a-, and b-type agents because d-type agents essentially act randomly and, thus, the d-type specific reduction is only of the order $0.5*n_d*p_a$ (i.e. $\approx 0.4\%$ for $p_a=0.05$ and $1.5\%$ for $p_a=0.2$). Fig. \ref{fig3a} reports the reduction $\Delta p_{te}$ with respect to the stationary state without audit $p_{te}^0$ \footnote{Note that these values have been determined as an average over $50$ time steps (steps $ 50 \to 100$) in the stationary state.}. This quantity, thus, measures by which percentage points tax evasion is reduced by the audit as compared to the non-audited society. It is obviously an increasing function of the audit probability $p_a$ and, as anticipated, also increases with the percentage of a-type agents. An interesting feature of the curves shown in Fig. \ref{fig3a} is the slope, which decreases with increasing $p_a$. We find that for small audit probabilities the percentage of detected tax evadors increases more strongly than for large $p_a$. Therefore, increasing the audit probability becomes more and more inefficient in increasing the number of detected tax evadors. A natural criterion for an efficient audit probability range is set by the requirement that the slope of the curves in Fig. \ref{fig3a} is larger than 'one'. This critical value (indicated by the circles in Fig. \ref{fig3a}) is reached at $p_a \approx 0.17$ for $0\%$ a-type agents and at $p_a \approx 0.23$ for $50\%$ a-type agents. Hence, subject to the aforementioned specifications an audit would in general be efficient for audit probabilities $p_a \lesssim 0.23$. 

\begin{figure}[h!] \begin{center}\includegraphics[width=12cm,clip=true]{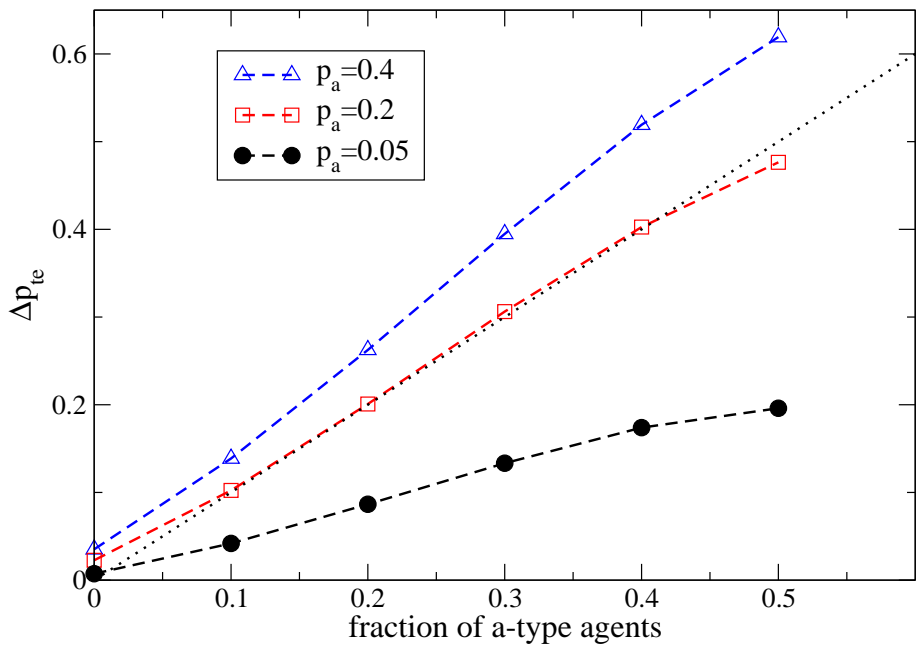}\end{center} \caption{Difference in tax evasion (stationary state) between zero and finite audit probability for audit probabilities $p_a=0.05, 0.2$ (see Fig. \ref{fig2}) and $p_a=0.4$.}
\label{fig3}
\end{figure}

Whereas our previous measure of efficiency is related to the effort which has to be put in an audit, efficiency can also be defined with regard to the success of an audit. In Fig. \ref{fig3} we plot again the difference of stationary tax evasion values $\Delta p_{te}$ with and without an audit, now as a function of the percentage of a-type agents. For example, given the percentage of a-type agents as $n_a=50\%$ we can read from Fig. \ref{fig3} that tax evasion is reduced by $\Delta p_{te}\approx 0.2$. As discussed above this reduction distributes approximately equal between the a- and b-type agents whereas the reduction of tax evading d-type agents is at most some few percentage points. If we define an audit as efficient when it can at least detect the percentage of a-type agents, then we see from Fig. \ref{fig3} that this criterion is almost fulfilled for $p_a=0.2$, but not for small audit probability $p_a=0.05$. In fact, let us consider the specific case of $n_a=50\%$ and audit probability $p_a=0.2$. Then we can read from Fig. \ref{fig2}b that the percentage of tax evasion is reduced from $\sim 91\%$ to $\sim 44\%$, i.e. $\Delta p_a\approx 0.47$. This reduction is composed of an audit induced reduction for a-type agents from $\sim 48\%$ to $\sim 25\%$ and for b-type agents from $\sim 31\%$ to $\sim 10\%$, while d-type agents contribute with a reduction from $\sim 8\%$ to $\sim 5\%$. Efficiency, in the sense defined here, thus means that the audit detects at least the number of tax evadors which corresponds to the fraction of inherently non-compliant tax payers (i.e. a-type agents). Yet, it does not mean that all the detected tax evadors are a-type agents because from the example above it is obvious that the reduction involves to approximately equal parts a- and b-type agents. 

Comparing these results with the results for the previously introduced definition of efficiency it turns out that an audit is efficient with regard to effort for $p_a \lesssim 0.23$, whereas, on the other hand, with regard to success an audit is efficient for $p_a \gtrsim 0.2$. Hence, for the society under consideration the optimal audit probability $p_a$ would be in the range of $0.2 \lesssim p_a \lesssim 0.23$, corresponding to the boundaries of both, otherwise disjunct, efficiency ranges.

Of course, other parameter values may lead to different ranges for the optimal audit probability. Moreover, in both cases audit efficiency is a pure quantity measure, whereas in the real world costs and revenue aspects may distort the two quantity measures. This notwithstanding, and subject to the model specifications, our results seem to indicate that tax audit rates applied in many countries, e.g. less than one percent in the U.S. according to \cite{bloom06}, may be suboptimally low. To this extent, our results also differ from those of \cite{zak09} or \cite{lz08}, who find that even small audit probabilities lead to tax compliance in the very long run.

\begin{figure}[thb]
\begin{center}
\includegraphics[width=12cm,clip=true]{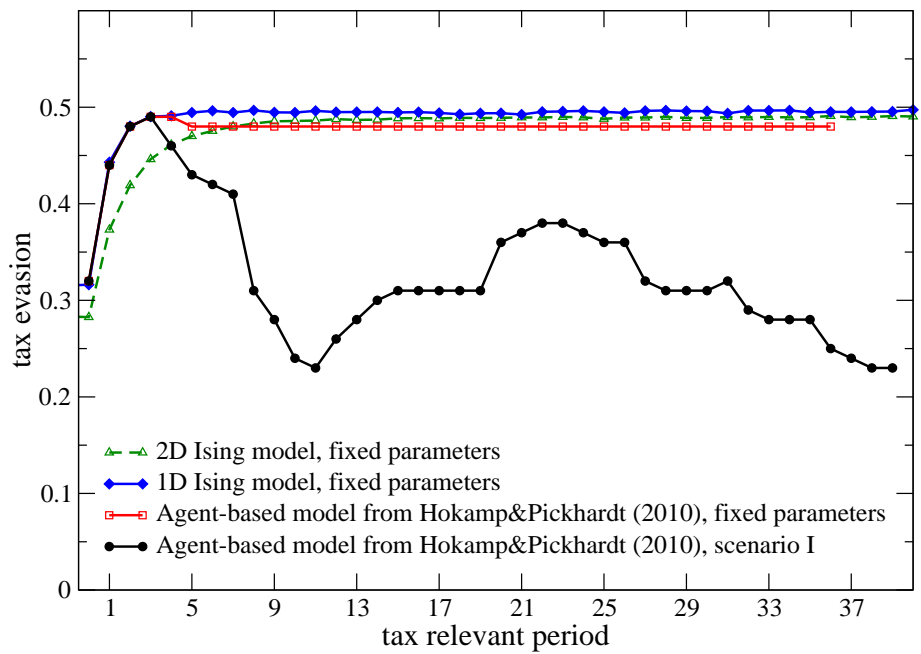}
\end{center}
\caption{Time evolution of tax evasion for (a) scenario I of the agent-based model by \cite{hopi10} [dots]; (b) agent-based model by \cite{hopi10}, but with fixed parameters [squares]; (c) Ising model on a square lattice with fixed parameters [triangles]; (d) Ising model on a ringworld with fixed parameters [diamonds].}
\label{fig6}
\end{figure}

\subsubsection{Reproducing Results from an Agent-based Economics Model}
\label{sec:HOPI}
Finally, we show in this section how the dynamics of the agent-based model by \textcolor{blue}{Hokamp and Pickhardt} (\textcolor{blue}{2010}) can be captured within the present multi-agent Ising-model approach. Solid dots in Fig. \ref{fig6} show the original dynamics in terms of percentage of tax evasion obtained for the first scenario and the set of governmental tax policy parameter changes used by \cite{hopi10}, Table 1, and Figure 1, first line from below. These tax policy parameters are the audit probability, tax rate, penalty rate and a parameter which specifies the complexity of the tax laws. Obviously, there is no direct mapping of these parameters to our Ising type approach since, for example, a monetary penalty rate presupposes the definition of a tax relevant income, whereas the Ising model can only deal with compliant or non-compliant tax payers.

However, we will demonstrate in this section that both approaches lead to very similar dynamics when the multi-agent Ising model is defined with appropriate field and temperature parameters for individual agents. Therefore, we restrict to a constant parameter set, i.e. the \cite{hopi10}, Table 1, tax policy parameters of the first tax period. For this fixed parameter set, the original results of the Hokamp and Pickhardt approach are displayed in Fig. \ref{fig6} by the curve with square symbols. Interestingly, these results are by and large in line with recent findings of \cite{kleven11}, who report evidence from a large scale tax audit field experiment in Denmark and find a tax evasion rate of $44.9\%$ for self-employment income.   

Yet, with respect to the replication it is worth noting that the Hokamp and Pickhardt ring world model is constructed such that four neighbors to either the left or right of an agent are taken into account. Unfortunately, this is not compatible with the symmetry of the Ising model which requires $J_{ij}=J_{ji}$, as in Eq. \ref{eq:ising}. Thus, for definiteness we fit the Hokamp and Pickhardt results with a one-dimensional Ising model (ring world with two neighbors) and, in addition, show differences to the results obtained with the two-dimensional Ising model (square lattice with four neighbors) and the same parameter set.

The  curves labeled with diamonds and triangles in Fig. \ref{fig6} show our replication results. The initial conditions are generated by just $50$ percent a-types and $50$ percent c-types.\footnote{Note that the type distribution in \cite{hopi10}, for the case to be replicated, is $50$ percent a-types, $35$ percent b-types, no c-types and $15$ percent d-types.} This is because in the first time step of the Hokamp and Pickhardt model b-types must use their default option (i.e. full compliance) and d-types are on average also compliant, so that the joint share of b-types and d-types is on average equivalent to a $50$ percent c-type share. A-type agents are specified by fields $-3.8 < B_i < -1$ and temperature $T_i = 7.4$, whereas c-type agents have $6< B_i < 12$ and $T_i=5.5$. As already noted, all units are measured with respect to $J = 1$. Equilibrium is reached after $10$ time steps so that time step $\#11$ corresponds to the first tax relevant time step or year shown in Fig. \ref{fig6}. As of the second tax relevant time step, we reduce the percentage of c-type agents to $15$ percent (i.e. the d-type share in Hokamp and Pickhardt)\footnote{We use this procedure, instead of the d-type definition of section \ref{sec:ModelMulti}, because in the Hokamp and Pickhardt model d-types may also randomly overpay their tax bill, which is not possible in our multi-agent Ising model.} and introduce instead a $35$ percent share of b-types, which are characterized by $B=0$ and $T = 0.1$. Hence, the ratio $J/T$ is large for b-types so that they copy the behavior of others in their neighborhood. The number of a-type agents remains unchanged. Thus, in the present multi-agent Ising-based model, a-type, b-type and c-type behavior patterns are reproduced with different local fields and temperatures, whereas d-type behavior is approximated on average by an equivalent percentage of c-type agents.

It turns out that during the first few time steps the Ising model multi-agent approach yields a fairly good match with the Hokamp and Pickhardt results because the lines with triangles, diamonds and squares (dots) closely match \footnote{Note that the parameter set for the one-dimensional Ising model (diamonds) have been fitted to match the \cite{hopi10} data (squares) over the first few time steps. Results for the two-dimensional Ising model (triangles) are based on the same parameter set. Of course, by using a different parameter set the two-dimensional Ising model results could be fitted to the Hokamp and Pickhardt data as well. }. However, in the large time limit the latter leads to a saturation at slightly smaller values. This is essentially due to several reasons with minor effects. In particular, differences in the stochastic processes, in the c-type versus d-type modeling, in the individual extent of tax evasion, in the penalties charged and in the lattice structures. 

Moreover, it is worth noting that differences with the square lattice replication (triangles in Fig. \ref{fig6}) are due to the fact that the b-type agents may copy the a-type behavior much faster in the ringworld. As the (standard) Ising model is based on the assumption that agents are influenced only by their direct neighbors, the one-dimensional Ising model (ringworld) reduces the size of the social network to two agents, whereas the social network size is four agents in both the two dimensional square lattice Ising model and the Hokamp and Pickhardt ringworld model. Hence, given the parameter setting, in the one-dimensional Ising ringworld it is more likely that b-type agents are influenced by their a-type neighbors. Thus, in each time step, tax evasion is higher in the Ising ringworld [diamonds] compared to the Ising square lattice world [triangles] (see Fig. \ref{fig6}), except in the long run. Regarding the Hokamp and Pickhardt ringworld, copying of tax evasion behavior is comparatively fast because the average tax evasion in the social network is taken into account in every time step, which almost guarantees that tax evasion behavior quickly spreads.

\section{Concluding Remarks}
\label{sec:Con}

In this paper we have independently reproduced results obtained from an agent-based econophysics model on tax evasion. Next, we extended the standard econophysics model of tax evasion in such a way that different behavioral agent types can be included. This represents another novelty because it extends previous work of \cite{lz08} towards a society which is build up from four different types of agents. We then used the extended model for an analysis of audit efficiency. It turned out that under the given circumstances substantially higher audit probabilities, in comparison to those that are effectively applied in most countries, would be desirable to curb tax evasion. Further, overall tax compliance of the extent displayed in Fig. \ref{fig2} is achieved by a non-monetary 'penalty', which consists of periods of enforced after-audit compliance. Hence, a non-monetary 'penalty' combined with copying behavior due to interactions among agents may be an effective substitute for the standard approach of charging monetary penalties.   Moreover, the new feature moves econophysics models of tax evasion closer to agent-based models of tax evasion that are found in the economics domain. Therefore, we were also able to use our extended econophysics model for reproducing results originally obtained from an agent-based tax evasion model of the economics domain. Again, this is the first time that such a reproduction has been achieved with an econophysics model.   

An important finding of this reproduction is that a micro foundation of individual behavior, expressed in terms of individual utility or payoff functions in models of the economics domain, is not needed for the aggregate results obtained by these models. Put differently, rational individual utility maximizing behavior can be reproduced on the aggregate level by so called zero intelligence agents, some stochastic changes in individual behavior patterns, and a certain degree of direct agent interaction that is controlled by the parameter 'temperature', which influences all agents in a nonrival manner like a public good (i.e. tax morale, massmedia, etc.).

Moreover, econophysics tax evasion models are based on a natural stochastic process derived from the Ising model of ferromagnetism. This is one of their major advantages in comparison with models from the economics domain in which either simple stochastic processes are used or unrecognized attempts are made that essentially try to reproduce elements of the Ising model. Hence, if econophysics models of tax evasion would be complemented with a micro foundation they might well make a fundamental contribution to the analysis of tax evasion. By extending the standard econophysics model of tax evasion in a way that various behavioral agent types could be incorporated, we made a first step towards a micro foundation of econophysics models. Of course, further steps are required, but these tasks delineate a future research agenda.

\end{document}